\documentclass[twocolumn,showpacs,superscriptaddress,preprintnumbers,nofootinbib,prd]{revtex4-1}
\usepackage{amsmath,amssymb,amsfonts,slashed}
\usepackage{graphicx}
\usepackage{graphics}
\usepackage{dcolumn}
\usepackage{epsfig,color}
\usepackage{bm}
\usepackage{verbatim}
\usepackage{epstopdf}

\begin{document}

\title{Threshold corrections of $\chi_{\rm c}(2P)$ and $\chi_{\rm b}(3P)$ states and $J/\psi \rho$ and $J/\psi \omega$ transitions of the $X(3872)$ in a coupled-channel model}

\author{J. Ferretti}
\affiliation{CAS Key Laboratory of Theoretical Physics, Institute of Theoretical Physics, Chinese Academy of Sciences, Beijing 100190, China}
\affiliation{Center for Theoretical Physics, Sloane Physics Laboratory, Yale University, New Haven, Connecticut 06520-8120, USA}
\author{E. Santopinto}
\affiliation{INFN, Sezione di Genova, Via Dodecaneso 33, 16146 Genova, Italy}

\begin{abstract}
We calculate the masses of $\chi_{\rm c}(2P)$ and $\chi_{\rm b}(3P)$ states with threshold corrections in a coupled-channel model. Here, the meson quarkonium core is augmented by higher Fock components due to pair-creation effects.
According to our results, we interpret the resonances characterized by very small threshold corrections, like $\chi_{\rm b}(3P)$'s, as almost pure quarkonia, and those states characterized by non-negligible threshold corrections, like the $X(3872)$, as quarkonium cores plus meson-meson components. 
We also study the $J/\psi \rho$ and $J/\psi \omega$ hidden-flavor strong decays of the $X(3872)$.
The decays are calculated as the dissociation of one of these components ($D^0 \bar D^{0*}$) into a $c \bar c$ state ($J/\psi$) plus a light meson ($\rho$ or $\omega$) in a potential model. 
In particular, our result for the ratio between the $X(3872) \rightarrow J/\psi \omega$ and $J/\psi \rho$ widths (0.6) is compatible with the present experimental data ($0.8\pm0.3$) within the experimental error.
\end{abstract}

\maketitle

\section{Introduction}
In the latest years, several new quarkonium-like states, the so-called $XYZ$ mesons, have been included in the PDG \cite{Nakamura:2010zzi}.
Some of them can be interpreted as standard $c \bar c$ or $b \bar b$ mesons, namely they can be described in terms of a $Q \bar Q$ valence component only. Others show unusual properties, and thus cannot be accommodated within a standard quarkonium picture; in some cases, the situation is even more complicated, e.g. when only the experimental mass of the state is provided, without the possible quantum numbers and decay modes.
Some examples include the $X(3872)$ \cite{Choi:2003ue,Acosta:2003zx,Abazov:2004kp}, $X(4140)$ \cite{Aaltonen} and $X(4260)$ \cite{Choi:2007wga}.
The previous states are labeled as suspected exotics and interpreted as compact tetraquark states \cite{Jaffe:1976ih,Barbour:1979qi,Weinstein:1983gd,SilvestreBrac:1993ss,Brink:1998as,Maiani:2004vq,Barnea:2006sd,Santopinto:2006my,Ebert:2008wm,Deng:2014gqa,Zhao:2014qva,Anwar:2017toa}, meson-meson molecules \cite{Weinstein:1990gu,Manohar:1992nd,Tornqvist:1993ng,Martins:1994hd,Swanson:2003tb,Hanhart:2007yq,Thomas:2008ja,Baru:2011rs,Valderrama:2012jv,Aceti:2012cb,Guo:2013sya}, the result of kinematic or threshold effects caused by virtual particles \cite{Heikkila:1983wd,Pennington:2007xr,Li:2009ad,Danilkin:2010cc,Ortega:2012rs,charmonium,Ferretti:2013vua,Achasov:2015oia,Kang:2016jxw,Lu:2016mbb}, hadro-quarkonia (hadro-charmonia) \cite{Dubynskiy:2008mq,Guo:2008zg,Wang:2009hi,Voloshin:2013dpa,Li:2013ssa,Wang:2013kra,Brambilla:2015rqa,Alberti:2016dru,Panteleeva:2018ijz,Ferretti:2018kzy}, or the rescattering effects arising by anomalous triangular singularities \cite{Guo:2014iya,Szczepaniak:2015eza,Liu:2015taa}; see also Ref. \cite{Guo:2008yz}, where the $X(4260)$, with $1^{--}$ quantum numbers, was interpreted as a member of the first hybrid supermultiplet.
For a review, see Refs. \cite{Esposito:2016noz,Olsen:2017bmm,Guo:2017jvc}.

The quark structure of the $X(3872)$, discovered by Belle in $B$ meson decays \cite{Choi:2003ue}, is still an open puzzle.
This resonance has $1^{++}$ quantum numbers, a very narrow width, and a mass far below quark model (QM) predictions \cite{Nakamura:2010zzi,Eichten:1978tg,Godfrey:1985xj,Barnes:2005pb,Suzuki:2005ha,Li:2009zu}. 
This is why the pure charmonium interpretation of the $X(3872)$ as a $\chi_{c1}(2^3P_1)$ state is incompatible with the present experimental data. 
However, the charmonium picture provides precise estimations for other observables, suggesting that the wave function of the $X(3872)$ may contain a non-negligible charmonium component \cite{charmonium02}.
Because of this, a possible solution to the $X(3872)$ puzzle is to describe the meson as a $c \bar c$ core plus continuum effects, i.e. as a mixture between charmonium and molecular-type components \cite{Pennington:2007xr,Danilkin:2010cc,charmonium,charmonium02}.
The weights of the $c \bar c$ and meson-meson components in the wave function of the $X(3872)$ vary widely, according to the particular model one considers \cite{Ortega:2012rs,charmonium02,CRvB}. 

The $\chi_{\rm b}(3P)$ system was discovered by ATLAS in 2012 \cite{Aad:2011ih} and later confirmed by D0 \cite{Abazov:2012gh}. 
The two collaborations gave an estimate of the multiplet mass barycenter, but they did not provide results for the mass splittings between the members of the multiplet. 
It is worth noting that in Ref. \cite{Abazov:2012gh} the authors stated: {\it``Further analysis is underway to determine whether this structure is due to the $\chi_b(3P)$ system or some exotic bottom-quark state".}
Indeed, $\chi_b(3P)$ resonances are not very far from the first open-bottom decay thresholds. Thus, in principle, their wave functions may include non-negligible continuum components \cite{Ferretti:2013vua,Lu:2016mbb,Karliner:2014lta}. 
The previous exotic interpretations seem to be in contrast with the recent CMS results for the masses of the $\chi_{\rm b1}(3P)$, $10513.42\pm0.41\pm0.18$ MeV, and $\chi_{\rm b2}(3P)$, $10524.02\pm0.57\pm0.18$ MeV \cite{Sirunyan:2018dff}.

In this paper, we discuss the interpretation of quarkonium-like exotics as $Q \bar Q$ cores plus (higher Fock or molecular-like) meson-meson components. 
The starting point is the spectrum.
Here, differently from the case of Refs. \cite{charmonium,Ferretti:2013vua}, we do not perform a global fit to the quarkonium spectrum. We use a coupled-channel model, where the quarkonium core is augmented by higher Fock components due to virtual particles, and extract the net threshold corrections within a multiplet after a certain zero-mode energy is subtracted. 
The latter is just the smallest self-energy correction (in terms of absolute value) of a multiplet member.
According to the previous procedure, we can interpret the resonances with zero or very small net threshold correction, like $\chi_{\rm b}(3P)$'s, as pure or almost pure quarkonia, while the states with non-negligible net threshold correction, like the $X(3872)$, as quarkonium cores plus meson-meson components.

We also study the hidden-flavor $J/\psi \rho$ and $J/\psi \omega$ transitions of the $X(3872)$.
The $J/\psi \rho$ and $J/\psi \omega$ decay amplitudes are computed by combining a coupled-channel formalism \cite{charmonium,Ferretti:2013vua,charmonium02,bottomonium,Bijker:2009up} with the nonrelativistic diagrammatic approach to meson-meson scattering of Refs. \cite{Barnes-Swanson,Barnes-Swanson02}.
Specifically, the formalism of Refs. \cite{charmonium,Ferretti:2013vua,charmonium02,bottomonium,Bijker:2009up} is used to include meson-meson (or continuum) components, like $D \bar D^*$, in the pure $c \bar c$ wave function of the $\chi_{c1}(2^3P_1)$. Then, the dissociation matrix elements of the $D^0 \bar D^{0*}$ component into $J/\psi$ plus $\rho$ or $\omega$ are calculated as the low-energy scattering and quark recombination of $D^0$ and $\bar D^{0*}$ mesons in a potential model \cite{Barnes-Swanson,Barnes-Swanson02}.
Finally, our results are compared with the existing experimental data \cite{Nakamura:2010zzi}. 
It is worth noting that our result for the ratio between the $X(3872) \rightarrow J/\psi \omega$ and $X(3872) \rightarrow J/\psi \rho$ widths, i.e. 0.6, is is compatible with the present experimental data, i.e. $0.8\pm0.3$ \cite{Nakamura:2010zzi,delAmoSanchez:2010jr}, within the experimental error.

\section{Threshold mass-shifts in $\chi_{\rm c}(2P)$ and $\chi_{\rm b}(3P)$ multiplets}
\subsection{Threshold mass-shifts in a coupled-channel model}
\label{Threshold mass-shifts}
We study the properties of $\chi_{\rm c}(2P)$ and $\chi_{\rm b}(3P)$ quarkonium-like states by means of a coupled-channel model, where the quarkonium core is augmented by higher Fock components due to virtual particle effects.
To leading order in pair creation, the quarkonium-like meson wave function can be written as \cite{Heikkila:1983wd,charmonium,Ferretti:2013vua,charmonium02,bottomonium,Bijker:2009up}
\begin{equation}	
	\label{eqn:Psi-A}
	\begin{array}{l}	
	\left| \psi_A \right\rangle = {\cal N} \left[ \left| A \right\rangle + \displaystyle \sum_{BC \ell J} \int q^2 dq
	\left| BC q \ell J \right\rangle \frac{ \left\langle BC q \ell J \right| T^{\dagger} \left| A \right\rangle}{M_a - E_b - E_c} \right] ~. 
	\end{array}
\end{equation}
Here, $\left| \psi_A \right\rangle$ is made up of a zeroth order quark-antiquark configuration, $\left| A \right\rangle$, plus a sum over all the possible higher Fock components, $\left| BC \right\rangle$, due to the creation of quark-antiquark pairs with vacuum quantum numbers. 
The sum is extended over a complete set of intermediate meson-meson states, $\left| BC \right\rangle$, with energies $E_{b,c} = \sqrt{M_{b,c}^2 + q^2}$; $M_a$ is the physical mass of the meson $A$; $q$ and $\ell$ are the relative radial momentum and orbital angular momentum of $B$ and $C$, and $J$ is the total angular momentum, with $\bf{J} = \bf{J}_b + \bf{J}_c + \bm{\ell}$.

Below, we discuss a coupled-channel model to compute the physical masses of quarkonium-like mesons, $M_a$, with threshold corrections.
This is done under the following hypotheses: 
a) Only the closest complete set of accessible SU(N)$_{\rm flavor} \otimes$ SU(2)$_{\rm spin}$ open-flavor meson-meson intermediate states (e.g. $1S1S$, $1S1P$ or $1S2S$) can influence the multiplet structure\footnote{In the case of quarkonium states around the opening of the first open-flavor decay thresholds, the closest complete set of open-flavor intermediate states is easy to identify. For example, in the case of the $\chi_{\rm c}(2P)$ multiplet, we consider a complete set of $1S1S$ open-flavor intermediate-state mesons. They include $D \bar D$, $D \bar D^*$, $D^* \bar D^*$, $D_{\rm s} \bar D_{\rm s}$, $D_{\rm s} \bar D_{\rm s}^*$, $D_{\rm s}^* \bar D_{\rm s}^*$, $\eta_{\rm c} \eta_{\rm c}$, $\eta_{\rm c} J/\psi$, and $J/\psi J/\psi$; see Table \ref{tab:components}. If one deals with higher-lying quarkonium states, the identification of the relevant intermediate states may represent a more subtle task. This is because at higher energies several open-flavor decay thresholds, made up of meson-meson pairs with different combinations of angular momenta and principal quantum numbers, lie closer to one another.}. 
The other (lower or upper) meson-meson thresholds, which are further in energy, are supposed to give some kind of global or background contribution, which can be subtracted\footnote{This is an attempt to solve one of the main problems of the unquenched quark model \cite{Heikkila:1983wd,Pennington:2007xr,Li:2009ad,Danilkin:2010cc,Ortega:2012rs,charmonium,Ferretti:2013vua,Achasov:2015oia,Kang:2016jxw,Lu:2016mbb}, its well-known lack of convergence. Indeed, as the tower of meson-meson intermediate states is enlarged, the contribution of continuum or sea components to hadron observables keeps growing indefinitely.}; b) The presence of a certain complete set of open-flavor intermediate states does not affect the properties of a single resonance, but it influences those of all the multiplet members.
Thus, the net effect of the intermediate states on a quarkonium-like meson multiplet is similar to that of a spin-orbit or hyperfine splitting.

Given this, the physical masses of the multiplet members are computed as
\begin{equation}
	\label{eqn:new-Ma}
	M_a = E_a + \Sigma(M_a) + \Delta \mbox{ }.
\end{equation}
Here, $E_a$ is the bare mass of meson $A$, whose value is extracted from the relativized QM predictions of Ref. \cite{Godfrey:1985xj}.
\begin{equation}
	\label{eqn:self-a}
	\Sigma(M_a) = \sum_{BC} \int_0^{\infty} q^2 dq \mbox{ } 
	\frac{\left| \left\langle BC q \ell J \right| T^\dag \left| A \right\rangle \right|^2}{M_a - E_b - E_c}  
\end{equation}
is a self-energy correction, where the sum is extended over a complete set of intermediate meson-meson states $BC$. 
$\Delta$ is the smallest self-energy correction (in terms of absolute value) of the multiplet (see Secs. \ref{chiC(2P)} and \ref{ChiB}).
The model parameters, which we need in the calculation of the $\left\langle BC q \ell J \right| T^\dag \left| A \right\rangle$ vertices, were fitted to the open-flavor strong decays of charmonia \cite[Table II]{charmonium} and bottomonia \cite[Table I]{Ferretti:2013vua}; see also \cite[Table II]{Ferretti:2015rsa}.
Therefore, for each multiplet, there is only one free parameter, $\Delta$.

\subsection{$\chi_{\rm c}(2P)$ multiplet}
\label{chiC(2P)}
We calculate the threshold mass shifts of the $\chi_{\rm c}(2P)$ multiplet members due to a complete set of ground state $1S 1S$ meson loops, like $D \bar D$, $D \bar D^*$, and so on.
\begin{figure}[htbp] 
\centering 
\includegraphics[width=6cm]{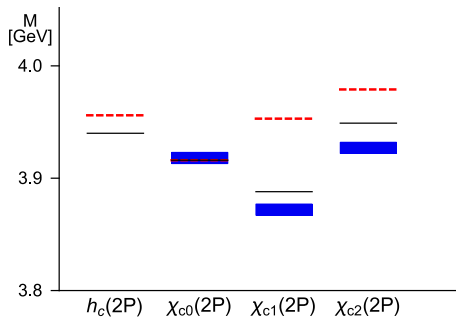}
\caption{Masses of the $\chi_{\rm c}(2P)$ multiplet members with threshold corrections. Blue boxes, dashed and continuous lines stand for experimental \cite{Nakamura:2010zzi}, calculated bare and physical masses, respectively.} 
\label{fig:threshold-ChiC(2P)}
\end{figure}

The values of the bare masses, $E_a$, are extracted from the relativized model \cite{Godfrey:1985xj}, those of the physical masses, $M_a$, from the PDG \cite{Nakamura:2010zzi}, with the exception of $h_{\rm c}(2P)$. 
As the latter state is still unobserved, for its physical mass we use the same value as the bare one \cite{Godfrey:1985xj}.
The self-energy corrections, $\Sigma(M_a)$, are computed as discussed in Sec. \ref{Threshold mass-shifts}, using the model parameter values of Refs. \cite{charmonium,charmonium02}; we get: $\Sigma(M_{h_{\rm c}(2P)}) = -119$ MeV, $\Sigma(M_{\chi_{\rm c0}(2P)}) = -103$ MeV, $\Sigma(M_{\chi_{\rm c1}(2P)}) = -168$ MeV, $\Sigma(M_{\chi_{\rm c2}(2P)}) = -133$ MeV. Thus, $\Delta = \Sigma(M_{\chi_{\rm c0}(2P)})$. Finally, in Table \ref{tab:ChiC(2P)-splittings} we report the values of the calculated physical masses of the $\chi_{\rm c}(2P)$ multiplet members. See also Figure \ref{fig:threshold-ChiC(2P)}.

It is worth noting that: I) Our theoretical predictions agree with the data within the typical error of a QM calculation, of the order of $30-50$ MeV; II) Among the $\chi_{\rm c}(2P)$ multiplet members, the $\chi_{\rm c1}(2P)$ receives the largest contribution from the continuum. This non-negligible continuum contribution, resulting in a large self-energy correction, is necessary to lower the relativized QM prediction, 3.95 GeV, towards the observed value of the mass, 3871.69 MeV \cite{Nakamura:2010zzi}; III) In the $\chi_{\rm c}(2P)$ case, threshold effects break the usual mass pattern of a $\chi$-type multiplet, namely $M_{\chi_0} < M_{\chi_1} \approx M_{\rm h} < M_{\chi_2}$. 
\begin{table}[htbp]
\small
\centering
\begin{tabular}{ccccc} 
\hline 
\hline
State                        & $E_a$             & $\Sigma(M_a) - \Delta$           & $M_a^{\rm th}$           & $M_a^{\rm exp}$ \\
                                &             [MeV]  &                                       [MeV] &  [MeV]                        &  [MeV] \\
\hline
$h_{\rm c}(2P)$       & 3956               & $-16$                                      & 3940                            & -- \\
$\chi_{\rm c0}(2P)$ & 3916               & 0                                             & 3916                            & 3918  \\
$\chi_{\rm c1}(2P)$ & 3953               & $-65$                                      & 3888                            & 3872  \\
$\chi_{\rm c2}(2P)$ & 3979               & $-30$                                      & 3949                            & 3927  \\                          
\hline                                
$h_{\rm b}(3P)$       & 10541            & $-4$                                         & 10538                          & 10519$^\dag$ \\
$\chi_{\rm b0}(3P)$ & 10522             & 0                                              & 10522                          & 10500$^\dag$  \\
$\chi_{\rm b1}(3P)$ & 10538             & $-2$                                         & 10537                          & 10512  \\
$\chi_{\rm b2}(3P)$ & 10550             & $-7$                                         & 10543                          & 10528$^\dag$  \\                             
\hline 
\hline
\end{tabular}
\caption{Comparison between the experimental masses \cite{Nakamura:2010zzi} of $\chi_{\rm c}(2P)$ and $\chi_{\rm b}(3P)$ states and theoretical predictions, as explained in the text. The bare masses, $E_a$, are extracted from Refs. \cite{Godfrey:1985xj,Godfrey:2015dia}. The experimental results denoted by $\dag$ are extracted from Ref. \cite{Godfrey:2015dia}, where the authors used predicted multiplet mass splittings in combination with the measured $\chi_{\rm b1}(3P)$ mass.}
\label{tab:ChiC(2P)-splittings}  
\end{table}

\begin{table*}
\footnotesize
\begin{tabular}{cccccccccccccc} 
\hline 
\hline
State               & $J^{PC}$ & $B \bar B$ & $B\bar B^*, \bar BB^*$ & $B^*\bar B^*$ & $B_s \bar B_s$ & $B_s \bar B_s^*, \bar B_s B_s^*$ & $B_s^* \bar B_s^*$ &  $B_c \bar B_c$ & $B_c \bar B_c^*, \bar B_c B_c^*$ & $B_c^* \bar B_c^*$ &  $\eta_b \eta_b$ & $\eta_b \Upsilon$ & $\Upsilon \Upsilon$ \\ \hline
$h_{\rm b}(3^1P_1)$ & $1^{+-}$ & -- & $\ell = 0,2$ & $\ell = 0,2$ & -- & $\ell = 0,2$ & $\ell = 0,2$ & -- & $\ell = 0,2$ & $\ell = 0,2$ & -- & $\ell = 0,2$ & -- \\                             
$\chi_{\rm b0}(3^3P_0)$ & $0^{++}$ & $\ell = 0$ & -- & $\ell = 0,2$ & $\ell = 0$ & -- & $\ell = 0,2$ &  $\ell = 0$ & -- & $\ell = 0,2$ & $\ell = 0$ & -- & $\ell = 0,2$ \\
$\chi_{\rm b1}(3^3P_1)$ & $1^{++}$ &-- & $\ell = 0,2$ & $\ell = 2$ & -- & $\ell = 0,2$ & $\ell = 2$ &  -- & $\ell = 0,2$ & $\ell = 2$ & -- & -- & $\ell = 2$ \\
$\chi_{b2}(3^3P_2)$ & $2^{++}$ & $\ell = 2$ & $\ell = 2$ & $\ell = 0,2$ & $\ell = 2$ & $\ell = 2$ & $\ell = 0,2$ & $\ell = 2$ & $\ell = 2$ & $\ell = 0,2$ & $\ell = 2$ & -- & $\ell = 0,2$ \\  
\hline
State               & $J^{PC}$ & $D \bar D$ & $D\bar D^*, \bar DD^*$ & $D^*\bar D^*$ & $D_s \bar D_s$ & $D_s \bar D_s^*, \bar D_s D_s^*$ & $D_s^* \bar D_s^*$ & $\eta_{\rm c} \eta_{\rm c}$ & $\eta_{\rm c} J/\psi$ & $J/\psi J/\psi$ & & &   \\ \hline
$h_{\rm c}(2^1P_1)$ & $1^{+-}$ & -- & $\ell = 0,2$ & $\ell = 0,2$ & -- & $\ell = 0,2$ & $\ell = 0,2$ & -- & $\ell = 0,2$ & -- & & & \\                       
$\chi_{\rm c0}(2^3P_0)$ & $0^{++}$ & $\ell = 0$ & -- & $\ell = 0,2$ & $\ell = 0$ & -- & $\ell = 0,2$ & $\ell = 0$ & -- & $\ell = 0,2$ & & &   \\
$\chi_{\rm c1}(2^3P_1)$ & $1^{++}$ &-- & $\ell = 0,2$ & $\ell = 2$ & -- & $\ell = 0,2$ & $\ell = 2$ &  -- & -- & $\ell = 2$ & & &   \\
$\chi_{c2}(2^3P_2)$ & $2^{++}$ & $\ell = 2$ & $\ell = 2$ & $\ell = 0,2$ & $\ell = 2$ & $\ell = 2$ & $\ell = 0,2$ & $\ell = 2$ & -- & $\ell = 0,2$  & & &   \\
\hline 
\hline
\end{tabular}
\caption{Meson-meson higher Fock components, $\left| BC \right\rangle$, in heavy-quarkonium-like meson wave functions, $\left| \psi_{\rm A} \right\rangle$. The relative orbital angular momentum between $B$ and $C$, $\ell$, is also shown. The symbol -- means that the $A \rightarrow BC$ coupling is forbidden by selection rules. In the case of $c \bar c - c \bar c$ and $b \bar b - b \bar b$ intermediate states, like $\eta_{\rm c} \eta_{\rm c}$ or $\eta_{\rm b} \eta_{\rm b}$, the quantum number selection rules also include $G$-parity conservation. 
For the definition of pseudoscalar-vector states, like $D^\pm D^{*\mp}$, we follow the prescription of Refs. \cite{Tornqvist:1993ng,Swanson:2003tb,Thomas:2008ja}. Specifically, we write the pseudoscalar-vector molecular-type states as linear combinations of $P \bar V$ and $\bar P V$ components with T\"ornqvist's relative sign convention, namely $C \left| P \bar V \pm \bar P V \right\rangle = \pm \left| P \bar V \pm \bar P V \right\rangle$.} 
\label{tab:components} 
\end{table*}

\subsection{$\chi_{\rm b}(3P)$ multiplet}
\label{ChiB}
We calculate the mass shifts within the $\chi_{\rm b}(3P)$ multiplet due to $1S 1S$ meson loops. 
The values of the physical masses are extracted from Refs. \cite{Nakamura:2010zzi,Godfrey:2015dia}, those of the bare masses from Ref. \cite{Godfrey:2015dia}. 
The self-energy corrections are calculated with the model parameter values of Ref. \cite{Ferretti:2013vua}; we get: $\Sigma(M_{h_{\rm b}(3P)}) = -116$ MeV, $\Sigma(M_{\chi_{\rm b0}(3P)}) = \Delta = -112$ MeV, $\Sigma(M_{\chi_{\rm b1}(3P)}) = -114$ MeV, $\Sigma(M_{\chi_{\rm b2}(3P)}) = -119$ MeV. Finally, in Table \ref{tab:ChiC(2P)-splittings} we report the values of the calculated physical masses of the $\chi_{\rm b}(3P)$ multiplet members and compare them with the experimental data.

According to our results: I) The threshold effects are supposed to be negligible and compatible with zero in the $\chi_{\rm b}(3P)$ case. Because of this, we interpret $\chi_{\rm b}(3P)$ states as (almost) pure bottomonia; II) Unlike the $\chi_{\rm c}(2P)$ case, the usual mass pattern within a $\chi$-type multiplet, namely $M_{\chi_0} < M_{\chi_1} \approx M_{\rm h} < M_{\chi_2}$, is now respected. 

Our predictions are in agreement with those of Refs. \cite{Godfrey:1985xj,Godfrey:2015dia} within the error of a QM calculation.
On the contrary, the results of Refs. \cite{Ferretti:2013vua,Lu:2016mbb} and the phenomenological predictions of Ref. \cite{Karliner:2014lta} would suggest the presence of non-negligible mixing effects between $\chi_{\rm b}(3P)$ states and continuum or molecular-type components. 
Recent experimental data from CMS \cite{Sirunyan:2018dff}, $M_{\chi_{\rm b1}(3P)} = 10513.42\pm0.41\pm0.18$ MeV and $M_{\chi_{\rm b2}(3P)} = 10524.02\pm0.57\pm0.18$ MeV, seem to agree with the present results for the mass patterns within the $\chi_{\rm b}(3P)$ multiplet.

\section{Continuum components in a coupled-channel model}
\label{Continuum components}
The operator $T^\dagger$ of Eq. (\ref{eqn:Psi-A}) induces a continuum component in an initially pure valence state $\left| A \right\rangle$ \cite{bottomonium}.
The procedure to calculate the norm of this continuum component is discussed in the following.

We define the ratio
\begin{equation}
	\label{eqn:Ri}
	\mathcal R_i = \frac{\Sigma(M_i) - \Delta}{\Sigma(M_i)}  \mbox{ },
\end{equation}
where $\Sigma(M_i)$ is the self-energy correction of Eq. (\ref{eqn:self-a}), the free parameter $\Delta$ is defined in Eq. (\ref{eqn:new-Ma}), and the index $i = 1, ..., N_{\rm mult}$ identifies each member of the meson multiplet with $N_{\rm mult}$ components. For example, in the $\chi_{\rm c}(2P)$ case we have $N_{\rm mult} = 4$, with $1 \rightarrow h_{\rm c}(2P)$, ..., $4 \rightarrow \chi_{\rm c2}(2P)$. For each member of the multiplet, we also define an effective and renormalized pair-creation strength
\begin{equation}
	\label{eqn:gamma0i-ren}
	\tilde \gamma_{0,i}^{\rm eff} = \gamma_0^{\rm eff} \sqrt{\mathcal R_i}  \mbox{ },
\end{equation}
where the effective pair-creation strength, $\gamma_0^{\rm eff}$, is that of \cite[Eq. (12)]{bottomonium}.
After doing this, we can calculate the norm of the continuum (or molecular-type) component for each member of the multiplet via \cite{bottomonium} 
\begin{equation}
	\label{eqn:Pa-sea}
	P_a^{\rm sea} = \sum_{BC\ell J} \int_0^\infty q^2 dq \mbox{ } 
	\frac{\left| \left\langle BC q  \, \ell J \right| T^\dag \left| A \right\rangle \right|^2}{(M_a - E_b - E_c)^2}  \mbox{ },
\end{equation}
where we replace the effective pair-creation strength, $\gamma_0^{\rm eff}$, in the vertex, $\left\langle BC q  \, \ell J \right| T^\dag \left| A \right\rangle$, with the effective and renormalized pair-creation strengths of Eq. (\ref{eqn:gamma0i-ren}), $\tilde \gamma_{0,i}^{\rm eff}$. For example, in the $X(3872)$ case $\tilde \gamma_{0,X(3872)} = 0.623 \mbox{ } \gamma_0$.

Finally, as an example, we provide results for the normalization of the continuum and valence components of the $X(3872)$ in Table \ref{tab:continuum}.
\begin{table}[htbp]
\small
\centering
\begin{tabular}{ccc}
\hline 
\hline
State                                                     & Component        & Probability \\
\hline
$X(3872)$ or $\chi_{\rm c1}(2^3P_1)$ & $D \bar D^*$                              & 0.816 \\
                                                              & $D^* \bar D^*$                          & 0.028  \\
                                                              & $D_{\rm s} \bar D_{\rm s}^*$     & 0.005  \\
                                                              & $D_{\rm s}^* \bar D_{\rm s}^*$  & 0.003  \\                                                             
                                                              & $J/\psi J/\psi$                            & $\approx 0$  \\  
\hline                                                              
                                                              & TOT sea (continuum)                          & 0.853  \\    
                                                              & valence                                      & 0.147  \\                                                                                                                                                                                                                                                                                                                                                                                                                                                                                                   
\hline 
\hline
\end{tabular}
\caption{Probability to find the $X(3872)$ or $\chi_{\rm c1}(2^3P_1)$ in its valence, $P_a^{\rm val}$, or meson-meson continuum component, $P_a^{\rm sea}$, calculated via Eq. (\ref{eqn:Pa-sea}). The two probabilities are related by $P_a^{\rm val} = 1 - P_a^{\rm sea}$.}  
\label{tab:continuum}  
\end{table}

\section{$J/\psi \rho$ and $J/\psi \omega$ hidden-flavor transitions of the $X(3872)$ in a coupled-channel model} 
\label{transitions in the UQM}
We calculate $J/\psi \rho$ and $J/\psi \omega$ hidden-flavor transitions of the $X(3872)$ in a coupled-channel model \cite{charmonium,Ferretti:2013vua,bottomonium,Bijker:2009up}. 

The transitions can be seen as two-step processes.
At a first stage, the $c \bar c$ $\chi_{c1}(2^3P_1)$ meson is ``dressed" with open-charm meson-meson continuum components, like $D \bar D$, $D \bar D^*$, and so on \cite[Table VIII and Eqs. (14)]{charmonium02}. This is done by means of the formalism of Refs. \cite{charmonium,Ferretti:2013vua,charmonium02,bottomonium}. 
The wave function of the $X(3872)$ is thus made up of a $\chi_{c1}(2^3P_1)$ core -- indicated as $\left| A \right\rangle$ in \cite[Eq. (12)]{charmonium02} -- plus open-charm $D \bar D$, $D \bar D^*$, ... higher Fock components -- indicated as $\left| BC \right\rangle$ in \cite[Eq. (12)]{charmonium02}. 
At a second stage, the $D^0 \bar D^{0*}$ continuum component of the $X(3872)$ dissociates into a $c \bar c$ meson, $J/\psi$, and a light one, $\rho$ or $\omega$ -- indicated as $\left| D \right\rangle$ and $\left| E \right\rangle$ in Eqs. (\ref{eqn:cross-section}) and (\ref{eqn:loop}).
The $BC \rightarrow DE$ scattering (or dissociation) amplitudes are computed with the non-relativistic potential model formalism of Refs. \cite{Barnes-Swanson,Barnes-Swanson02} and Sec. \ref{diagrammatic} where, for simplicity, we consider a single harmonic oscillator parameter, $\alpha = 0.5$ GeV, for the wave functions of the five mesons, $A$, $B$, $C$, $D$ and $E$.

\subsection{Diagrammatic approach to meson-meson scattering}
\label{diagrammatic}
In this section, we briefly remind the diagrammatic approach to meson-meson scattering of Refs. 
\cite{Barnes-Swanson,Barnes-Swanson02}.
\begin{figure}
\centering
\includegraphics[width=7cm]{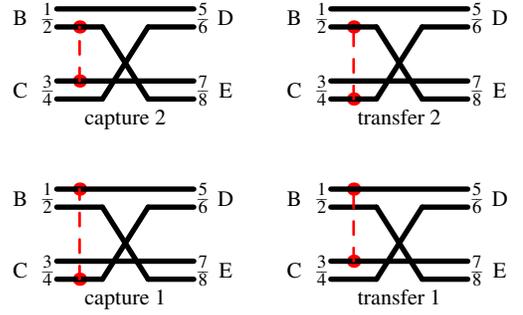}
\caption{The four diagrams which contribute to the scattering amplitude \cite{Barnes-Swanson,Barnes-Swanson02}. The numbers 1, 2, 3 and 4 indicate the valence quarks (antiquarks) before recombination; the labels 5, 6, 7 and 8 indicate the valence quarks after the recombination process has taken place.} 
\label{fig:4diagrams}
\end{figure} 
The Hamiltonian of the model is\footnote{The Hamiltonians used to compute the bare meson masses, \cite[Eqs. (2)]{Godfrey:1985xj}, and the meson-meson scattering amplitudes, Eq. (\ref{eqn:Hmodel}) and \cite[Eqs. (1) and (2)]{Barnes-Swanson}, do not coincide. Our choice is motivated by the fact that the relativized quark model \cite{Godfrey:1985xj} is characterized by a very complicated Hamiltonian; its usage would make the calculation of the scattering amplitudes of Fig. \ref{fig:4diagrams} a tough task. While in principle it is true that one should use the same quark-quark interaction both to bind mesons and calculate their scattering amplitudes, a more rigorous calculation is out of the scope of the present paper.}
\begin{equation}
   \label{eqn:Hmodel}	
   \begin{array}{l}
   	H = H_0 + H_{\rm I} = \displaystyle \sum_i - \frac{\hbar^2}{2 m_i} \nabla^2_i + \displaystyle \sum_{i < j} H_{ij}  \mbox{ },
   \end{array}
\end{equation}
where
\begin{equation}
	\label{eqn:H-ij}
	\begin{array}{l}
		H_{ij} = \frac{\lambda_i^a}{2} \frac{\lambda_j^a}{2} \left[ \frac{\alpha_s}{r_{ij}} - 
		\frac{3 \beta}{4} \mbox{ } r_{ij}  - \frac{8 \pi \alpha_s}{3 m_i m_j} \mbox{ } {\bf S}_i \cdot {\bf S}_j \delta({\bf r}_{ij}) \right] \mbox{ },
	\end{array} 
\end{equation}
$\lambda_i^a$ and $\lambda_j^a$ are Gell-Mann color matrices and $r_{ij}$ is the relative coordinate between the quarks $i$ and $j$; see Fig. \ref{fig:4diagrams}.
The matrix elements of the interacting Hamiltonian are given by
\begin{equation}
	\label{eqn:scatteringME}
	\left\langle D E \right| H_{\rm I} \left| BC \right\rangle = \delta({\bf P}_{\rm i} - {\bf P}_{\rm f}) \mbox{ } 
	h_{\rm fi} \mbox{ },
\end{equation}
where ${\bf P}_{\rm i} = {\bf P}_b + {\bf P}_c$ and ${\bf P}_{\rm f} = {\bf P}_d + {\bf P}_e$ are initial and final momenta of the meson-meson pair, respectively. 
There are four diagrams which contribute to the $BC \rightarrow DE$ scattering amplitude (see Fig. \ref{fig:4diagrams}).
The matrix elements $h_{\rm fi}$ of a particular diagram can be written as the product of five contributions
\begin{equation}
	h_{\rm fi} = S I_{\rm flavor} I_{\rm color} I_{\rm spin} I_{\rm space} \mbox{ },
\end{equation}
where $S=-1$, resulting from the permutation of fermion operators in the scattering matrix elements, is the ``signature" phase. $I_{\rm flavor}$, $I_{\rm color}$ and $I_{\rm spin}$ are flavor, color and spin matrix elements, respectively, and $I_{\rm space}$ the spatial matrix element of the potential of Eq. (\ref{eqn:H-ij}) \cite{Barnes-Swanson}.

As an example, we calculate the spatial matrix element of the spin-spin potential,
\begin{equation}
	\label{eqn:Vss}
	\begin{array}{rcl}
	V_{\rm ss} & = & - \frac{8 \pi \alpha_{\rm s}}{3 m_i m_j} {\bf S}_i \cdot {\bf S}_j \frac{1}{(2\pi)^3} \int d^3r_{ij} e^{i {\bf p}_{ij} \cdot 
	{\bf r}_{ij}} 	\delta({\bf r}_{ij}) \\
	& = & - \frac{1}{(2\pi)^3} \frac{8 \pi \alpha_{\rm s}}{3 m_i m_j} {\bf S}_i \cdot {\bf S}_j 
	\end{array} \mbox{ },	
\end{equation}
in the case of the capture 1 (c1) diagram of Fig. \ref{fig:4diagrams}; one has \cite{Barnes-Swanson}
\begin{equation}
	\label{eqn:Ic1ss}
	\begin{array}{rcl}
	I_{\rm c1, ss} & = & - \frac{1}{(2\pi)^3} \frac{8 \pi \alpha_s}{3 m_1 m_4} {\bf S}_1 \cdot {\bf S}_4 
	\int d^3 k' \mbox{ } \Phi_d^*\left({\bf k}' - \frac{1}{2} {\bf P}_d\right) \\ 
	& \times & \int d^3k \mbox{ } \Phi_e^*\left({\bf k} - \frac{1}{2} {\bf P}_b\right) \Phi_b\left({\bf k} + \frac{1}{2}{\bf P}_d\right) 
	\Phi_c\left({\bf k} - \frac{1}{2}{\bf P}_d\right)  \\
	& = & - \frac{1}{(2\pi)^3} \frac{8 \pi \alpha_s}{3 m_1 m_4} {\bf S}_1 \cdot {\bf S}_4 \mbox{ }
	(2 \pi)^{3/2} \mbox{ } \Psi_d^*(0) \int d^3k \mbox{ } \\
	& \times & \Phi_e^*\left({\bf k} - \frac{1}{2} {\bf P}_b\right) \Phi_b\left({\bf k} + \frac{1}{2}{\bf P}_d\right) 
	\Phi_c\left({\bf k} - \frac{1}{2}{\bf P}_d\right) \mbox{ }.
	\end{array}	
\end{equation}
The matrix elements of Eq. (\ref{eqn:scatteringME}) are calculated by taking ${\bf P}_d$ along the $\hat z$ axis.
Finally, the $B C \rightarrow DE$ cross section is given by:
\begin{equation}
	\label{eqn:cross-section}
	\sigma_{bc \rightarrow de} = \frac{(2\pi)^4 E_b E_c E_d E_e}{E_{bc}^2} \mbox{ } 
	\frac{\left|{\bf P}_d\right|}{\left|{\bf P}_b\right|}  \int d\Omega_b \left|h_{\rm fi}\right|^2  \mbox{ }.
\end{equation}

\subsection{Results}
Analogously to what is done in the positronium case \cite{Itzykson:1980rh}, the decay amplitude is written as
\begin{equation}
\label{eqn:loop}
	\begin{array}{l}
		\Gamma = \left[\int P_b^2dP_b \mbox{ } \frac{\left|\left\langle A| T^\dag |BC \right\rangle\right|^2}
		{(M_a - E_{bc})^2 + \frac{\Gamma_a^2}{4}}  \left| {\bf v}_b - {\bf v}_c \right| 
		\sigma_{bc \rightarrow de}  \right]  \left|\Psi_{BC}(0)\right|^2    
	\end{array}	\mbox{ },
\end{equation}
where the term in square brackets is the convolution product between a distribution function, describing the probability to find the $\left| BC\right\rangle = \left| D^0 \bar D^{0*} \right\rangle$ component in the wave function of $\left| A\right\rangle = \left| X(3872) \right\rangle$ \cite[Table VIII]{charmonium02}, $\left| {\bf v}_b - {\bf v}_c \right|$ is the difference between the velocities of the mesons $B$ and $C$, and $\frac{\Gamma_a^2}{4}$ [where $\Gamma_a$ is the experimental total width of the $X(3872)$] in the denominator is necessary for the convergence of the calculation, the mass of the $X(3872)$ being very close to the $D^0 \bar D^{0*}$ threshold \cite{charmonium02}.
We also have to multiply the amplitude by $\left|\Psi_{BC}(0)\right|^2$, which is the squared wave function of the $BC$ molecule evaluated in the origin, namely its probability density.
The harmonic oscillator parameter of $\Psi_{BC}({\bf r})$, $\alpha_{bc}$, is determined from the condition
\begin{equation}
	\int d^3r \mbox{ } \Psi_{BC}^*({\bf r}) \mbox{ } r \mbox{ } \Psi_{BC}({\bf r}) = \left\langle r \right\rangle_{X(3872)}^{\rm molecule}
	\mbox{ };
\end{equation}
here, $\left\langle r \right\rangle_{X(3872)}^{\rm molecule} \simeq \frac{1}{\sqrt{2\mu {\mathcal E}_{\rm b}}} \simeq 10$ fm is the dimension of the $X(3872)$ in the $D^0 \bar D^{0*}$ molecular interpretation, where $\mu$ is the $D^0 \bar D^{0*}$ reduces mass and ${\mathcal E}_{\rm b}$ its binding energy.
To get results, the width of Eq. (\ref{eqn:loop}) has to be integrated over the Breit-Wigner mass distribution of the $\rho$ and $\omega$ mesons, because $M_{J/\psi} + M_{\rho/\omega} > M_{D^0} + M_{\bar D^{0*}}$.

Finally, the results of our calculation are:
\begin{equation}
	\Gamma_{X(3872) \rightarrow J/\psi \rho} = 10 \mbox{ keV }
\end{equation}
and
\begin{equation}
	\Gamma_{X(3872) \rightarrow J/\psi \omega} = 6 \mbox{ keV }.
\end{equation}
It is worth noting that our result for the ratio between the $X(3872) \rightarrow J/\psi \omega$ and $X(3872) \rightarrow J/\psi \rho$ amplitudes,
\begin{equation}
	\label{eqn:ratio-th}
	\frac{\Gamma^{\rm th}_{X(3872) \rightarrow J/\psi \omega}}{\Gamma^{\rm th}_{X(3872) \rightarrow J/\psi \rho}} 
	= 0.6 \mbox{ },
\end{equation}
is compatible with the present experimental data \cite{Nakamura:2010zzi,delAmoSanchez:2010jr},
\begin{equation}
	\label{eqn:ratio-exp}
	\frac{\Gamma^{\rm exp}_{X(3872) \rightarrow J/\psi \omega}}{\Gamma^{\rm exp}_{X(3872) \rightarrow J/\psi \rho}} 
	= 0.8\pm0.3  \mbox{ },
\end{equation}
within the experimental error.
See also Table \ref{tab:ratio}, where the result of Eq. (\ref{eqn:ratio-th}) is compared to those of other calculations.	

\begin{table}[htbp]  
\small
\caption{Our theoretical result for the ratio between the amplitudes $X(3872) \rightarrow J/\psi \omega$ and $X(3872) \rightarrow J/\psi \rho$ is compared to those of other studies and the experimental data \cite{Nakamura:2010zzi}.}
\begin{center}
\begin{tabular}{cc} 
\hline 
Source                              & Value of $\frac{\Gamma_{X(3872) \rightarrow J/\psi \omega}}{\Gamma_{X(3872) \rightarrow J/\psi \rho}}$  \\ 
\hline 
Experiment \cite{Nakamura:2010zzi}  & $0.8\pm0.3$   \\
Ref. \cite{Suzuki:2005ha}           & $\approx 2$   \\
Ref. \cite{Meng:2007cx}             & $1.0 \pm 0.3$ \\
Ref. \cite{Li:2012cs}               & 0.42 \\
Ref. \cite{Takeuchi:2014rsa}        & $1.27 - 2.24$ \\
Ref. \cite{Gamermann:2009uq}        & 1.4 \\
Ref. \cite{Zhou:2017txt}        & $0.58 - 0.92$ \\
Present work                        & 0.6           \\    
\hline
\end{tabular}
\end{center}
\label{tab:ratio}  
\end{table}

In Ref. \cite{Suzuki:2005ha}, the author estimated the ratio between the $X(3872) \rightarrow J/\psi \omega$ and $J/\psi \rho$ amplitudes, $R_{\omega/\rho} \approx 2$, in a semi-quantitative way, in which both $J/\psi \rho$ and $J/\psi \omega$ are produced through $D$- and $\bar D^*$-exchange between the $D \bar D^*$ pair.
In Ref. \cite{Meng:2007cx}, the ratio between the $X(3872) \rightarrow J/\psi \rho$ and $J/\psi \omega$ widths was calculated by using the re-scattering mechanism and effective Lagrangians, based on the chiral symmetry and heavy quark symmetry. The final result for the ratio is $R_{\rho/\omega} = 1.0 \pm 0.3$.
In Ref. \cite{Li:2012cs}, the description of the $X(3872)$ as a $D^0 \bar D^{0*}$ molecule is analyzed within the framework of both one-pion-exchange and one-boson-exchange models and the result for the ratio between the $X(3872) \rightarrow J/\psi \mbox{ } 3\pi$ and $J/\psi \mbox{ } 2\pi$ widths is 0.42.
In Ref. \cite{Takeuchi:2014rsa}, the authors calculated the ratio between the $X(3872) \rightarrow J/\psi \mbox{ } 3\pi$ and $J/\psi \mbox{ } 2\pi$ widths, $R_{\omega/\rho} = 1.27 - 2.24$, in a $c \bar c$-two-meson hybrid model, where the two-meson state consists of the $D^0 \bar D^{0*}$, $D^+ \bar D^{-*}$, $J/\psi \rho$ and $J/\psi \omega$ channels.

\section{Conclusion}
We calculated the masses of $\chi_{\rm c}(2P)$ and $\chi_{\rm b}(3P)$ states with threshold corrections in a coupled-channel model. In our approach, the quarkonium core of the mesons of interest is augmented by higher Fock components due to pair-creation effects \cite{charmonium,Ferretti:2013vua,bottomonium}. The pair-creation mechanism is inserted at the quark level and the one-loop diagrams are computed by summing over a complete set of accessible SU(N)$_{\rm f}$ $\otimes$ SU(2)$_{\rm s}$ ground-state mesons.
In order to calculate the mass shifts within a certain meson multiplet, we made the following hypotheses: a) Only the closest complete set of meson-meson open-flavor intermediate states can influence the multiplet structure. The other (lower or upper) meson-meson thresholds, which are further in energy, are supposed to give some kind of global or background contribution, which can be subtracted; b) The presence of a complete set of meson-meson open-flavor intermediate states does not affect the properties of a single resonance, but it influences the behavior of all the multiplet members.
In our interpretation, the $\chi_{\rm c0}(2P)$ is pure charmonium, while the $\chi_{\rm c1}(2P)$ and $\chi_{\rm c2}(2P)$ are characterized by a relevant threshold component; the $\chi_{\rm b}(3P)$ states are (almost) pure bottomonia.
The possible importance of coupled-channel effects within a certain quarkonium multiplet depends on the energy distance between the meson-meson thresholds and the masses of the multiplet members.
For example, in the $\chi_{\rm c}(2P)$ multiplet, the $X(3872) - D \bar D^*$ energy difference is of the order of 1 MeV, while in the $\chi_{\rm b}(3P)$ case the closest open-flavor thresholds are a few tens of MeV away. 

We also calculated the $X(3872) \rightarrow J/\psi \rho$ and $J/\psi \omega$ hidden-flavor strong decays.
The transitions can be seen as two-step processes.
At a first stage, the $c \bar c$ core of the $X(3872)$ is ``dressed" with open-charm meson-meson continuum components, like $D \bar D$, $D \bar D^*$, and so on \cite{charmonium,charmonium02,bottomonium,Ferretti:2013vua}.  
At a second stage, the $D^0 \bar D^{0*}$ continuum component of the $X(3872)$ dissociates into a $c \bar c$ meson, $J/\psi$, and a light one, $\rho$ or $\omega$.
The dissociation amplitude is computed with the non-relativistic potential model formalism of Refs. \cite{Barnes-Swanson,Barnes-Swanson02} and Sec. \ref{diagrammatic}. 
It is worth observing that our result for the ratio between $J/\psi \rho$ and $J/\psi \omega$ widths, Eq. (\ref{eqn:ratio-th}), is compatible with the corresponding experimental data, Eq. (\ref{eqn:ratio-exp}), within the experimental error. 

In conclusion, the results displayed here and in Refs. \cite{charmonium,charmonium02} suggest the interpretation of the $X(3872)$ as the superposition of a $c \bar c$ core and meson-meson continuum (or molecular-type) components, like $D \bar D^*$ and so on.
On the contrary, the present results are compatible with a ``pure" bottomonium interpretation for $\chi_{\rm b}(3P)$ states.

\begin{acknowledgments}
This work is supported in part by the Sino-German Collaborative Research Center ``Symmetries and the Emergence of Structure in QCD" (NSFC Grant No. 11621131001, DFG Grant No. TRR110), and by the NSFC (Grant No. 11747601).
\end{acknowledgments}

\end{document}